\documentclass[sigconf]{acmart}

\setcopyright{none}
\acmDOI{}
\acmISBN{}
\acmConference[]{}{}{} 
\acmYear{}

\renewcommand\footnotetextcopyrightpermission[1]{}

\hypersetup{
    pdftitle={ContextBranch: Conversation Branching for Exploratory LLM Interactions},
    pdfauthor={Bhargav Chickmagalur Nanjundappa, Spandan Maaheshwari},
    pdfsubject={Software Engineering, AI, Human-Computer Interaction},
    pdfkeywords={Large Language Models, Conversation Management, Version Control, 
                 Software Engineering, Context Isolation, Human-AI Interaction}
}

\usepackage{balance}
\usepackage{graphicx}
\usepackage{xcolor}
\usepackage{booktabs}
\usepackage{algorithm}
\usepackage{algpseudocode}
\algrenewcommand\alglinenumber[1]{\scriptsize #1:}
\usepackage{enumitem}
\usepackage{listings}
\usepackage{subcaption}
\usepackage{hyperref}
\usepackage[T1]{fontenc}
\usepackage{tikz}
\usetikzlibrary{fit,positioning}
\usepackage{amsthm}
\usepackage{threeparttable}

\newtheorem{property}{Property}

\newcommand{\system}{ContextBranch}

\begin{document}

\title{Context Branching for LLM Conversations: A Version Control Approach to Exploratory Programming}

\author{Bhargav Chickmagalur Nanjundappa}
\affiliation{
  \institution{Northeastern University}
  \city{Boston}
  \country{USA}
}
\email{chickmagalurnanjun.b@northeastern.edu}

\author{Spandan Maaheshwari}
\affiliation{
  \institution{Northeastern University}
  \city{Boston}
  \country{USA}
}
\email{maaheshwari.s@northeastern.edu}

\begin{abstract}
Large Language Models (LLMs) have become integral to software engineering workflows, yet their effectiveness degrades significantly in multi-turn conversations. Recent studies demonstrate an average 39\% performance drop when instructions are delivered across multiple turns, with models making premature assumptions and failing to course-correct (Laban et al., 2025). This degradation is particularly problematic in exploratory programming tasks where developers need to investigate alternative approaches without committing to a single path. Current solutions force users into a false dichotomy: continue in a context-polluted conversation where the LLM becomes increasingly confused, or start fresh and lose all accumulated context.

We present ContextBranch, a conversation management system that applies version control semantics to LLM interactions. ContextBranch provides four core primitives—checkpoint, branch, switch, and inject—enabling users to capture conversation state, explore alternatives in isolation, and selectively merge insights. We evaluate ContextBranch through a controlled experiment with 30 software engineering scenarios featuring intentionally polluting explorations. Branched conversations achieved 2.5\% higher overall response quality compared to linear conversations (p=0.010, Cohen's d=0.73), with large improvements in focus (+4.6\%, d=0.80) and context awareness (+6.8\%, d=0.87). Benefits were concentrated in complex scenarios involving conceptually distant explorations, with peak improvements of 13.2\%. Branching reduced context size by 58.1\% (31.0→13.0 messages), eliminating irrelevant exploratory content. Our work establishes conversation branching as a fundamental primitive for AI-assisted exploratory work, demonstrating that isolation prevents context pollution when exploring alternatives.
\end{abstract}


\sloppy
\maketitle

\section{Introduction}

Large Language Models have become integral to software development, yet their effectiveness degrades significantly in extended conversations. While benchmarks focus on single-turn tasks, real-world programming assistance involves multi-turn dialogues where developers iteratively refine requirements and explore alternatives. Recent analysis of over 200,000 conversations across 15 production LLMs reveals a 39\% average performance drop when instructions are delivered across multiple turns~\cite{laban2025lost}. Models make premature assumptions, over-rely on earlier responses, and fail to course-correct---when they take a wrong turn, they become lost and do not recover.

This problem is particularly acute during exploratory programming. Consider a developer optimizing a data pipeline who has established context about their Python implementation, performance constraints, and business requirements over several turns. They now want to explore: \emph{``What if we used Rust for the bottleneck sections?''} Current interfaces force an unsatisfactory choice of continuing the conversation which pollutes the Python-focused context, starting  fresh on the other hand loses all established context, and managing multiple windows creates cognitive overhead with no systematic synthesis path.

Software engineers solved an analogous problem decades ago with version control. Git enables developers to checkpoint state, branch for isolated experimentation, and selectively merge changes. We argue that conversational AI interactions require similar primitives. Just as code exploration benefits from checkpoint-restore semantics, conversation exploration requires the ability to preserve state, branch into isolated contexts, and selectively reintegrate insights.

We present \system{}, a conversation management system that applies version control semantics to LLM interactions. \system{} provides four core primitives: \textbf{checkpoint} captures conversation state at decision points, \textbf{branch} creates isolated exploratory contexts, \textbf{switch} navigates between branches without cross-contamination, and \textbf{inject} selectively merges insights from experimental threads. The system maintains deterministic context states through content-addressable message storage and ensures branch isolation through controlled injection. We evaluate \system{} through a controlled experiment comparing branched and linear conversations across 30 software engineering scenarios to simulate realistic but distracting tangents.

\subsection{Contributions}

\begin{enumerate}[leftmargin=*, itemsep=1pt, topsep=2pt]
\item We formalize the conversation branching problem as a mismatch between linear interfaces and exploratory work patterns (§\ref{sec:related}).
\item We present \system{}'s design, including its branching model, checkpoint semantics, and isolation guarantees (§\ref{sec:approach}).

\item Through a controlled experiment with 30 scenarios featuring intentional context pollution, we demonstrate that conversation branching improves response quality by 2.5\% overall with large effects on focus and context awareness, while reducing context size by 58.1\%.

\item We release \system{} as open source with tools for reproducible conversation experiments.
\end{enumerate}

\section{Background and Motivation}

\subsection{Multi-Turn Conversation Degradation}

The effectiveness of LLMs in extended conversations has only recently received systematic study. Laban et al.~\cite{laban2025lost} conducted the most comprehensive analysis to date, simulating over 200,000 conversations across 15 production models including GPT-4.1, Gemini 2.5 Pro, and Claude 3.7 Sonnet. Their experimental design compared single-turn instruction delivery (where all requirements are specified upfront) against multi-turn delivery (where requirements emerge across multiple exchanges). The results reveal systematic degradation: models averaged 39\% lower performance in multi-turn scenarios, with some tasks showing drops exceeding 60\%. 

Error analysis identified three distinct failure modes. \emph{Premature assumption} occurs when models commit to interpretations before receiving complete information, often within the first few turns of conversation. \emph{Prior response anchoring} describes models over-weighting their previous outputs, making it difficult to introduce new perspectives or corrections. Most critically, \emph{correction failure} demonstrates that when models take a wrong conversational turn, they struggle to recover even when explicitly provided with corrective information. As the authors conclude, ``when LLMs take a wrong turn in a conversation, they get lost and do not recover''~\cite{laban2025lost}.

This multi-turn degradation compounds with known architectural limitations in how models process context. Liu et al.~\cite{liu2024lost} demonstrated that models exhibit U-shaped attention patterns across context windows, attending strongly to initial and final content while neglecting information in the middle---a phenomenon they term ``lost in the middle.'' Their experiments with document retrieval tasks showed performance degrading by up to 30\% when relevant information appeared in the middle third of the context window compared to the first or last third. For multi-turn conversations, this means earlier context receives disproportionate weight while mid-conversation details are effectively forgotten, even when those details are critical to the current query.

Chroma Research~\cite{chroma2025context} investigated a related phenomenon they call ``context rot.'' Their experiments revealed that irrelevant context imposes a dual-task burden: models must simultaneously retrieve relevant information and perform reasoning, degrading performance on both tasks. In controlled experiments, adding just 10\% irrelevant content to prompts reduced accuracy by 23\%. This finding has direct implications for exploratory conversations, where earlier discussion of rejected approaches becomes irrelevant context that pollutes subsequent reasoning.

\subsection{Impact on Software Engineering Workflows}

These limitations manifest acutely in AI-assisted programming. Barke et al.~\cite{barke2023grounded} surveyed 113 professional developers using conversational AI tools for software engineering tasks. Their findings reveal that 73\% of respondents identified lack of context retention as their primary frustration, describing scenarios where they were forced to ``repeatedly provide the same background information in multi-turn interactions.'' Developers also reported that LLMs frequently ``generate plausible but incorrect solutions'' when context becomes incomplete or polluted across conversation turns.

The problem is particularly severe during exploratory programming---the iterative process of investigating alternatives, evaluating trade-offs, and refining solutions that characterizes much of software development work. 

Consider a concrete scenario drawn from our preliminary observations of developer behavior. A senior engineer is optimizing a data processing pipeline that currently uses Python's Pandas library to process daily batch files of approximately 10GB. After establishing context about the current architecture, performance bottlenecks (I/O-bound operations taking 45 minutes), and business constraints (must complete within 1-hour maintenance window), the engineer has explored several optimization strategies across 12 conversational turns: improving Pandas operations, using Dask for parallel processing, and optimizing data schemas.

At this point, the engineer wonders whether a fundamentally different approach might be warranted: ``What if we rewrote the bottleneck sections in Rust and called them from Python?'' This question represents a significant architectural shift---from pure Python optimization to a hybrid polyglot approach. However, current conversational interfaces provide no good option for exploring this alternative:

\textbf{Option 1: Continue in the same conversation.} The LLM's response will be influenced by the entire 12-turn Python-focused discussion. The model may inappropriately anchor to previous Python-centric solutions, fail to consider Rust's distinct trade-offs, or conflate the two approaches. When the engineer later wants to return to pure Python optimization, the Rust discussion has polluted the context, making it impossible to get unbiased Python recommendations.

\textbf{Option 2: Start a new conversation.} The engineer must manually reconstruct all relevant context: the 10GB file size, 45-minute runtime, 1-hour constraint, data schema details, tried optimizations, and their results. This reconstruction is tedious, error-prone (critical details may be forgotten), and non-deterministic (rephrasing context produces different conversation trajectories).

\textbf{Option 3: Maintain multiple chat windows.} Opening a fresh window for Rust exploration preserves the original Python thread but creates cognitive overhead. The engineer must now mentally track which insights from the Rust conversation are worth incorporating back into the Python discussion. There is no systematic mechanism for synthesis, leading to either abandonment of valuable insights or manual copy-paste workflows that are clumsy and error-prone.

\subsection{The Case for Conversation Branching}

Software engineers solved a structurally analogous problem decades ago. Before version control systems, exploring alternative code implementations required either overwriting working code (risking loss of the original) or manually maintaining multiple file copies (creating synchronization nightmares). Version control systems like Git introduced checkpoint-and-branch semantics: developers can mark a known-good state, create isolated branches for experimentation, and selectively merge successful changes back to the main development line.

We argue that conversational AI interactions exhibit the same structural need for branching. The parallel is precise:

\begin{itemize}[leftmargin=*, itemsep=1pt]
\item \textbf{Code commits} $\leftrightarrow$ \textbf{Conversation checkpoints}: Both capture known-good states that can be restored
\item \textbf{Code branches} $\leftrightarrow$ \textbf{Conversation branches}: Both enable isolated exploration without contaminating the main line
\item \textbf{Merge commits} $\leftrightarrow$ \textbf{Insight injection}: Both selectively incorporate valuable changes while discarding failed experiments
\item \textbf{Branch comparison} $\leftrightarrow$ \textbf{Approach comparison}: Both enable objective evaluation of alternatives
\end{itemize}

However, conversation branching introduces unique challenges not present in code version control. Code merging can rely on syntactic conflict detection (e.g., diff3 algorithms), while conversation merging requires semantic coherence. A git branch knows nothing about its siblings, while conversation branches may need to reference each other's insights. Code commits are discrete units of change, while conversation turns form a continuous dialogue where context matters deeply.

These differences suggest that naive application of Git semantics to conversations will be insufficient. A conversation branching system requires careful design to maintain context coherence, support deterministic exploration, and enable selective insight reintegration while respecting the fundamentally different nature of conversational versus code artifacts.

\subsection{Design Requirements}

Based on the documented limitations of multi-turn conversations and the needs of exploratory programming workflows, we identify four core requirements for a conversation branching system:

\textbf{R1: Deterministic Checkpointing.} The system must capture complete conversation state such that restoring a checkpoint produces identical context for subsequent interactions. This enables reproducible exploration of alternatives.

\textbf{R2: Branch Isolation.} Operations on one branch must not affect other branches. This prevents context pollution and enables truly independent exploration of conflicting approaches.

\textbf{R3: Selective Reintegration.} The system must support cherry-picking specific insights from experimental branches back to the main thread, rather than requiring all-or-nothing merges.

\textbf{R4: Low Overhead.} Checkpoint and branch operations must add negligible latency to conversation flows to maintain interactive usability.

These requirements guide the design of \system{}, which we present in the following section.
\section{Related Work}
\label{sec:related}

We position \system{} within research on context management for LLMs, version control systems, and AI-assisted programming, highlighting the gaps our work addresses.

\textbf{Context Management in LLMs.} Techniques like chain-of-thought prompting~\cite{wei2022chain} and retrieval-augmented generation~\cite{lewis2020retrieval} improve reasoning and knowledge access but do not address multi-turn degradation within conversations. Memory systems for conversational agents (MemPrompt~\cite{madaan2022memory}, Reflexion~\cite{shinn2023reflexion}, MemGPT~\cite{packer2023memgpt}) maintain context across sessions through external memory and self-reflection, but focus on persistent memory rather than isolated exploration of alternatives. Tree of Thoughts~\cite{yao2023tree} enables multi-path reasoning through deliberate exploration of thought branches, achieving 74\% success on challenging reasoning tasks versus 4\% for standard prompting. However, ToT operates on reasoning steps within single queries, not extended conversational interactions. Dialogue state tracking systems~\cite{wu2019transferable} maintain structured conversation state but assume a single coherent dialogue path without branching capabilities. Recent surveys~\cite{yi2024survey} confirm that existing multi-turn dialogue systems lack mechanisms for parallel exploration and selective reintegration of insights.

\textbf{Version Control and Collaborative Systems.} Git~\cite{torvalds2010git} provides conceptual foundations for branching through content-addressable storage and lightweight pointers, but its merge algorithms rely on syntactic conflict detection unsuitable for conversational content. CRDTs~\cite{shapiro2011crdts} and operational transformation~\cite{ellis1989ot} enable concurrent editing with automatic conflict resolution, assuming the goal is convergence to a single shared state. Conversation branching requires different semantics: maintaining intentionally divergent exploratory threads with explicit selective injection rather than automatic merging. Local-first software principles~\cite{kleppmann2019local} emphasize user control over data and offline operation, values we adopt but apply to conversation state management rather than document editing.

\textbf{AI-Assisted Programming.} Empirical studies reveal significant challenges with current tools. Barke et al.~\cite{barke2023grounded} characterize distinct interaction modes (acceleration versus exploration) and find that developers struggle to debug AI-generated code. GitHub's research with 2,000+ developers found that 73\% reported Copilot helped them stay in flow, but context retention remains problematic~\cite{github2022copilot}. Vaithilingam et al.~\cite{vaithilingam2022expectation} found that Copilot did not improve task completion time in controlled studies, with participants spending significant effort understanding and debugging generated code. More recent studies confirm that ``lack of understanding context by AI assistant'' remains a primary frustration~\cite{sergeyuk2024using}, and tools like Cursor show productivity gains but persistent increases in technical debt~\cite{cursor2024impact}. These findings motivate our focus on conversation management infrastructure that existing AI coding tools lack.

\system{} addresses gaps at the intersection of these areas. Unlike context management techniques that optimize linear conversations, we enable branching and parallel exploration. Unlike version control systems designed for code, we handle conversational content with semantic coherence requirements. Unlike AI programming tools that maintain only linear history, we provide checkpoint, branch, switch, and inject primitives as fundamental building blocks. Our key insight is that exploratory programming with LLMs requires conversation management primitives analogous to version control but adapted to conversational semantics, enabling workflows impossible in existing systems: deterministic checkpointing (R1), branch isolation (R2), selective reintegration (R3), and low overhead (R4).

\section{Approach}
\label{sec:approach}

We present \system{}, a conversation management system that enables deterministic branching and selective reintegration of insights in LLM interactions. Our design addresses the four requirements established in Section 2: deterministic checkpointing (R1), branch isolation (R2), selective reintegration (R3), and low overhead (R4).

\subsection{Design Principles}

\system{} is built on four core principles that distinguish it from naive conversation copying or existing context management approaches:

\textbf{Isolation through state capture.} Rather than maintaining a single mutable conversation history, \system{} treats each checkpoint as an immutable snapshot. Branches created from a checkpoint share the checkpoint's history but maintain independent futures. Operations on one branch cannot affect another, preventing the context pollution documented by Laban et al.~\cite{laban2025lost}.

\textbf{Determinism through content addressing.} Checkpoints use content-addressable storage where identical conversation histories map to identical checkpoint identifiers. This ensures that restoring a checkpoint always produces the same starting state, enabling reproducible exploration of alternatives. This approach draws inspiration from Git's object model~\cite{torvalds2010git} but adapts it to conversational semantics.

\textbf{Selectivity through explicit injection.} Unlike CRDTs that automatically merge concurrent edits~\cite{shapiro2011crdts}, \system{} requires users to explicitly select which messages to inject from experimental branches. This preserves conversational coherence and prevents automatic merging of semantically conflicting content.

\textbf{Composability through hierarchical branching.} Branches can be created from other branches, forming a tree structure that mirrors the exploratory nature of programming workflows. Each branch maintains a clear lineage to its parent checkpoint, enabling complex exploration patterns while preserving deterministic restoration.

\subsection{Core Abstractions}

We formalize conversation branching through three key abstractions: conversation state, checkpoints, and branches.

\begin{definition}[Conversation State]
A conversation state $S$ is an ordered sequence of messages:
$$S = \langle m_1, m_2, \ldots, m_n \rangle$$
where each message $m_i = (\text{\textit{role}}_i, \text{\textit{content}}_i, \text{\textit{metadata}}_i)$ consists of a role $\text{\textit{role}}_i \in \{\text{user}, \text{assistant}, \text{system}\}$, textual content $\text{\textit{content}}_i$, and optional metadata $\text{\textit{metadata}}_i$ (timestamps, token counts, model identifiers).
\end{definition}

\begin{definition}[Checkpoint]
A checkpoint $C$ is an immutable snapshot of conversation state at a decision point:
$$C = (\text{\textit{id}}_C, \text{\textit{parent}}_C, S_C, t_C)$$
where $\text{\textit{id}}_C = \text{hash}(S_C)$ is a content-addressable identifier, $\text{\textit{parent}}_C$ references the parent checkpoint (or $\bot$ for the root), $S_C$ is the captured state, and $t_C$ is the creation timestamp.
\end{definition}

\begin{definition}[Branch]
A branch $B$ extends a checkpoint with new messages:
$$B = (\text{\textit{id}}_B, C_B, S_B, \text{\textit{active}}_B)$$
where $\text{\textit{id}}_B$ is a unique branch identifier, $C_B$ is the source checkpoint, $S_B = S_{C_B} \oplus \Delta_B$ represents the checkpoint state extended with branch-specific messages $\Delta_B$, and $\text{\textit{active}}_B$ indicates whether the branch is currently selected for interaction.
\end{definition}

The operator $\oplus$ denotes sequence concatenation: if $S_{C_B} = \langle m_1, \ldots, m_k \rangle$ and $\Delta_B = \langle m_{k+1}, \ldots, m_n \rangle$, then $S_B = \langle m_1, \ldots, m_n \rangle$.

\subsection{Operations and Algorithms}

\system{} provides four primitive operations that satisfy requirements R1-R4.

\subsubsection{Checkpoint Creation}

Algorithm~\ref{alg:checkpoint} creates a checkpoint from the current conversation state.

\begin{algorithm}[ht]
\caption{Create Checkpoint}\label{alg:checkpoint}
\begin{algorithmic}[1]
\Require Current conversation state $S$, branch $B$
\Ensure Checkpoint $C$ with $S_C = S_B$
\State $\text{\textit{id}}_C \gets \text{SHA256}(S_B)$ \Comment{Content-addressable ID}
\If{checkpoint with $\text{\textit{id}}_C$ exists}
    \State \Return existing checkpoint \Comment{Deduplication}
\EndIf
\State $\text{\textit{parent}}_C \gets C_B$ \Comment{Link to parent checkpoint}
\State Store $(S_B, \text{metadata})$ indexed by $\text{\textit{id}}_C$
\State $t_C \gets \text{current\_timestamp}()$
\State \Return $C = (\text{\textit{id}}_C, \text{\textit{parent}}_C, S_B, t_C)$
\end{algorithmic}
\end{algorithm}

\textbf{Satisfies R1 (Determinism):} The content-addressable identifier ensures that identical conversation states always produce the same checkpoint. Restoring a checkpoint loads the exact message sequence $S_C$, guaranteeing deterministic starting conditions for subsequent exploration.

\textbf{Satisfies R4 (Low Overhead):} Checkpoint creation is $O(n)$ where $n$ is the number of messages, dominated by the hash computation. Deduplication prevents storage of redundant checkpoints when users repeatedly checkpoint the same state.

\subsubsection{Branch Creation}

Algorithm~\ref{alg:branch} creates an isolated branch from a checkpoint.

\begin{algorithm}[ht]
\caption{Create Branch}\label{alg:branch}
\begin{algorithmic}[1]
\Require Checkpoint $C$, branch name $\text{name}$
\Ensure New branch $B'$ with $S_{B'} = S_C$
\State $\mathit{id}_{B^{\prime}} \gets \mathrm{UUID}()$ \Comment{Unique branch identifier}
\State $S_{B'} \gets$ deep copy of $S_C$ \Comment{Isolation through copying}
\State $\Delta_{B'} \gets \langle \rangle$ \Comment{Empty delta initially}
\State $C_{B'} \gets C$ \Comment{Record source checkpoint}
\State Store branch metadata: $(\text{name}, t_{\text{created}}, C_{B'})$
\State \Return $B' = (\text{\textit{id}}_{B'}, C_{B'}, S_{B'}, \text{false})$
\end{algorithmic}
\end{algorithm}

\textbf{Satisfies R2 (Isolation):} The deep copy of $S_C$ ensures that modifications to $S_{B'}$ do not affect the checkpoint or other branches. Each branch maintains its own independent message sequence.

\textbf{Satisfies R4 (Low Overhead):} Branch creation is $O(n)$ for copying $n$ messages. In practice, modern Python implementations use copy-on-write semantics for sequences, making the actual cost negligible until messages are added.

\subsubsection{Branch Switching}

Switching between branches (Algorithm~\ref{alg:switch}) changes the active context for LLM interactions.

\begin{algorithm}[ht]
\caption{Switch Branch}\label{alg:switch}
\begin{algorithmic}[1]
\Require Target branch $B_{\text{target}}$, current branch $B_{\text{current}}$
\Ensure $B_{\text{target}}$ becomes active branch
\State $\text{\textit{active}}_{B_{\text{current}}} \gets \text{false}$ \Comment{Deactivate current}
\State $\text{\textit{active}}_{B_{\text{target}}} \gets \text{true}$ \Comment{Activate target}
\State Load $S_{B_{\text{target}}}$ into LLM context \Comment{No cross-contamination}
\State \Return $B_{\text{target}}$
\end{algorithmic}
\end{algorithm}

\textbf{Satisfies R2 (Isolation):} Switching branches replaces the entire conversation state with $S_{B_{\text{target}}}$. No messages from $B_{\text{current}}$ persist in the new context, preventing cross-contamination.

\textbf{Satisfies R4 (Low Overhead):} Branch switching is $O(1)$ in metadata updates. Loading $S_{B_{\text{target}}}$ into the LLM API is necessary regardless of branching and adds no overhead beyond normal conversation continuation.

\subsubsection{Selective Injection}

Algorithm~\ref{alg:inject} injects selected messages from a source branch into a target branch while preserving conversational coherence.

\begin{algorithm}[ht]
\caption{Inject Messages}\label{alg:inject}
\begin{algorithmic}[1]
\Require Source branch $B_s$, target branch $B_t$, message indices $I = \{i_1, \ldots, i_k\}$
\Ensure Target branch $B_t$ updated with selected messages
\State $M \gets \{m_i \mid i \in I, m_i \in \Delta_{B_s}\}$ \Comment{Extract messages}
\For{$m \in M$}
    \If{$\text{role}(m) = \text{assistant}$}
        \State Validate $m$ does not conflict with $S_{B_t}$
    \EndIf
\EndFor
\State $\text{insertion\_point} \gets |S_{C_{B_t}}|$ \Comment{After checkpoint, before target delta}
\State $S_{B_t} \gets S_{B_t}[1:\text{insertion\_point}] \oplus M \oplus S_{B_t}[\text{insertion\_point}+1:]$
\State Update metadata: record injection source and timestamp
\State \Return updated $B_t$
\end{algorithmic}
\end{algorithm}

\textbf{Coherence validation} (line 4-5) ensures injected assistant messages do not contradict the target branch's context. For example, injecting ``Let's use Rust'' into a Python-focused branch would fail validation if the target has already committed to a Python-only solution.

\textbf{Satisfies R3 (Selectivity):} Users explicitly specify which messages to inject via indices $I$. Only selected insights transfer between branches, preventing wholesale merging of conflicting approaches.

\textbf{Satisfies R2 (Isolation):} Injection operates on independent copies. The source branch $B_s$ remains unchanged; only the target branch $B_t$ is modified.

\subsection{Correctness Properties}

We establish three key properties that \system{} guarantees.

\begin{property}[Branch Isolation]
For any branches $B_i, B_j$ where $i \neq j$, operations on $B_i$ do not modify $S_{B_j}$.
\end{property}

\begin{proof}
Each branch maintains an independent copy of its conversation state (Algorithm~\ref{alg:branch}, line 2). Operations that modify a branch (sending messages, injection) operate only on that branch's local state. The only shared data structure is the checkpoint store, which contains immutable checkpoints.
\end{proof}

\begin{property}[Checkpoint Determinism]
For any checkpoint $C$ with state $S_C$, restoring $C$ at any time $t$ produces identical state $S_C$.
\end{property}

\begin{proof}
Checkpoints are immutable once created (Definition 2). The content-addressable identifier $\text{\textit{id}}_C = \text{hash}(S_C)$ ensures that the same conversation state always maps to the same checkpoint. Since $S_C$ is stored immutably, retrieval always returns the original state.
\end{proof}

\begin{property}[Injection Safety]
Injecting messages $M$ from branch $B_s$ into branch $B_t$ preserves the conversational coherence of $B_t$.
\end{property}

\begin{proof}[Proof sketch]
Algorithm~\ref{alg:inject} validates that assistant messages in $M$ do not conflict with existing context in $S_{B_t}$ (line 4-5). User messages are always safe to inject as they represent new information. The insertion point (line 7) ensures injected messages appear after the shared checkpoint history but before target-specific messages, maintaining chronological and semantic coherence.
\end{proof}

\subsection{Example Walkthrough}

We illustrate \system{}'s operations through the data pipeline scenario from Section 2.

\textbf{Initial conversation (turns 1-12):} The developer establishes context about Python optimization (Pandas, 10GB files, 45-minute runtime). The conversation state is $S_0 = \langle m_1, \ldots, m_{24} \rangle$ where messages alternate between user and assistant (12 turns = 24 messages).

\textbf{Checkpoint creation:} At turn 12, the developer creates checkpoint $C_1$:
$$C_1 = (\text{hash}(S_0), \bot, S_0, t_1)$$
This captures the Python-focused discussion as a known-good state.

\textbf{Branch for Rust exploration:} The developer creates branch $B_{\text{rust}}$ from $C_1$:
$$B_{\text{rust}} = (\text{uuid}(), C_1, S_0, \text{true})$$
With $B_{\text{rust}}$ active, the developer asks: ``What if we used Rust for bottleneck sections?'' This adds messages $m_{25}$ (user question) and $m_{26}$ (assistant response about Rust FFI). After 3 turns exploring Rust, $S_{B_{\text{rust}}} = S_0 \oplus \langle m_{25}, m_{26}, m_{27}, m_{28}, m_{29}, m_{30} \rangle$.

\textbf{Return to main thread:} The developer creates branch $B_{\text{main}}$ from $C_1$ to continue pure Python optimization:
$$B_{\text{main}} = (\text{uuid}(), C_1, S_0, \text{false})$$
Switching to $B_{\text{main}}$ restores $S_0$ with no Rust contamination. The developer explores parallel processing with Dask, generating messages $m_{31}$ through $m_{36}$.

\textbf{Selective injection:} After evaluating both approaches, the developer decides Rust's FFI overhead isn't worth it, but Rust's insight about memory-mapped I/O (in $m_{28}$) is valuable. They inject $m_{28}$ into $B_{\text{main}}$:
$$S_{B_{\text{main}}} = S_0 \oplus \langle m_{28} \rangle \oplus \langle m_{31}, \ldots, m_{36} \rangle$$

The conversation tree now has structure:
\begin{center}
\begin{tikzpicture}[
  level distance=1.5cm,
  level 1/.style={sibling distance=4cm},
  level 2/.style={sibling distance=2cm}
]
\node {$C_1$ (turn 12)}
  child {node {$B_{\text{rust}}$: 3 Rust turns}}
  child {node {$B_{\text{main}}$: Dask + injected insight}};
\end{tikzpicture}
\end{center}

This workflow was impossible in linear conversation interfaces. The developer explored conflicting approaches (Python vs. Rust) without contamination, then selectively merged valuable insights—satisfying all four requirements R1-R4.


\section{Implementation}
\label{sec:implementation}

We implement \system{}, a lightweight Python SDK that implements checkpoint-based conversation branching for LLM applications. Our implementation consists of 900 lines of core SDK code with zero dependencies, making it easily integrable into existing systems.  The implementation directly realizes the algorithms from Section~\ref{sec:approach} while providing a clean API for integration with existing LLM workflows.

\subsection{Architecture}

\begin{figure}[t]
\centering
\includegraphics[width=\columnwidth, alt={ContextBranch architecture showing components of Core SDK}]{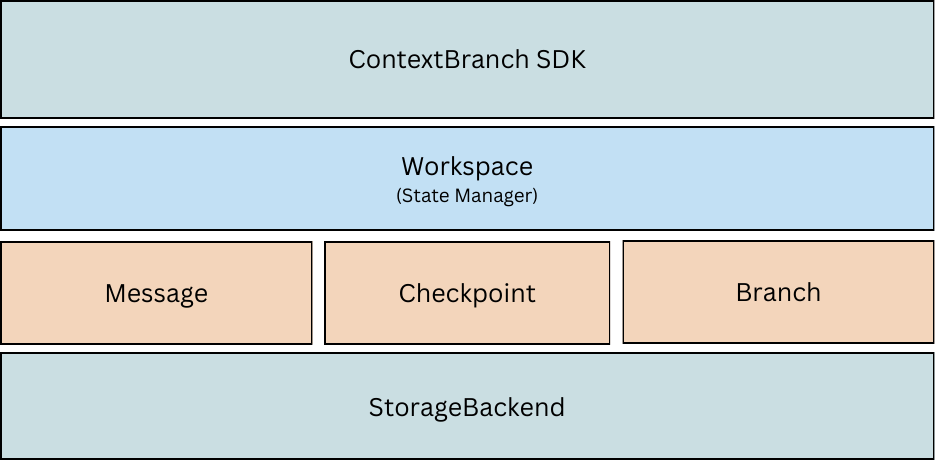}
\caption{\system{} Architecture}
\Description{ContextBranch architecture showing components of Core SDK}
\label{fig:architecture}
\end{figure}

\textbf{Core SDK.} The implementation is structured in three layers (Figure~\ref{fig:architecture}). The \textit{workspace layer} implements branch management and context isolation (\texttt{workspace.py}, 400 LOC; \texttt{branch.py}, 95 LOC), with deep-copy semantics ensuring R2 (branch isolation). The \textit{data layer} provides immutable message and checkpoint primitives with content-addressable identifiers (Algorithm 1 from Section~\ref{sec:approach} is implemented in \texttt{checkpoint.py}, 80 LOC).  The \textit{storage layer} abstracts persistence through a \texttt{StorageBackend} interface with file-based and in-memory implementations (\texttt{storage.py}, 180 LOC).

\textbf{Design Decisions.} We chose Python for rapid prototyping and broad LLM ecosystem compatibility. This minimizes integration friction and ensures the SDK works across diverse deployment environments. Storage backends use JSON for checkpoints (human-readable, debuggable) with SHA-256 hashing for deterministic checkpoint IDs (satisfying R1).

\subsection{API Design}

The public API exposes five core operations. Listing~\ref{lst:api_example} shows typical usage for a motivating scenario where the main conversation discusses the data pipeline requirements and explores different database options:

\begin{lstlisting}[language=Python, caption={ContextBranch API example: exploring database options with branch isolation.}, label={lst:api_example}, basicstyle=\small\ttfamily, frame=single]
from context_branching import ContextBranchingSDK, Message
# Initialize SDK with file-based persistence
sdk = ContextBranchingSDK(
    storage_backend="file",
    storage_path="./data"
)
workspace = sdk.create_workspace("db-decision")

# Main conversation: discuss pipeline requirements
workspace.add_message(Message(
    role="user",
    content="I need a database for 10M records, <1s query"
))
response = llm.chat_with_workspace(workspace, ...)
# ... continue main discussion ...

# Decision point: create checkpoint
cp_id = workspace.create_checkpoint("db-decision")

# Branch 1: Explore PostgreSQL
workspace.create_branch(cp_id, "explore-postgres")
workspace.switch_branch("explore-postgres")
workspace.add_message(Message(
    role="user", content="What about PostgreSQL?"
))
# ... LLM provides PostgreSQL analysis ...

# Branch 2: Explore MongoDB (from same checkpoint)
workspace.switch_branch("main")
workspace.create_branch(cp_id, "explore-mongo")
workspace.switch_branch("explore-mongo")
workspace.add_message(Message(
    role="user", content="What about MongoDB?"
))
# ... LLM provides MongoDB analysis ...

# Return to main, inject key insights
workspace.switch_branch("main")
workspace.inject_messages("explore-postgres", [0, 2])
workspace.inject_messages("explore-mongo", [0, 1])
# Continue with combined insights, no context pollution
\end{lstlisting}

\section{Evaluation}

We evaluate \system{} through a controlled experiment comparing branched conversations against traditional linear conversations across 30 realistic software engineering scenarios. Our evaluation addresses two questions: (1) Does conversation branching improve response quality when exploring alternatives? (2) Does branching reduce context pollution as measured by context size and response focus?

\subsection{Experimental Design}

\subsubsection{Scenario Construction}

We designed 10 software engineering problem types spanning common programming tasks: debugging concurrency issues, API design decisions, algorithm optimization, framework selection, database schema design, performance profiling, code refactoring, testing strategies, deployment architecture, and security implementations. Each problem type has three instantiations, yielding 30 scenarios total.

Each scenario includes five components: (1) \emph{context}—background about the system and constraints, (2) \emph{initial problem}—the developer's starting question, (3) \emph{exploration topics}—three alternative approaches to investigate, (4) \emph{final query}—a follow-up question requiring focus on the original problem, and (5) \emph{ground truth}—explicit lists of concepts the LLM should remember versus concepts from exploratory tangents that should not distract from the answer.

Table~\ref{tab:example-scenarios} shows three representative scenarios illustrating our design. Exploration topics are intentionally chosen to be conceptually distant from the core problem, simulating realistic but distracting tangents that developers naturally pursue during exploratory programming.

\begin{table}[ht]
\centering
\caption{Representative Scenarios Demonstrating Intentional Context Pollution}
\label{tab:example-scenarios}
\small
\begin{tabular}{p{1.8cm}p{2.5cm}p{2.9cm}}
\toprule
\textbf{Scenario} & \textbf{Core Problem} & \textbf{Polluting Explorations} \\
\midrule
Bug Debugging & Intermittent test failures in CI pipeline with PostgreSQL foreign key constraints & Jest→Vitest migration, Prettier/ESLint configuration, LaunchDarkly feature flags \\
\midrule
API Design & Choosing REST vs GraphQL for web, mobile, and IoT clients needing offline sync & Docker/Kubernetes containerization, OAuth2/JWT authentication, Prometheus/Grafana monitoring \\
\midrule
Performance & N+1 database queries causing 5-8s dashboard load with real-time update requirements & React vs Vue frontend rewrite, Next.js server-side rendering, TypeScript migration \\
\bottomrule
\end{tabular}
\end{table}

\subsubsection{Conversation Structure}

Each scenario follows a standardized three-phase structure designed to simulate realistic exploratory programming workflows:

\textbf{Phase 1: Context establishment} (5-6 turns). The developer and LLM establish the core problem, system constraints, and relevant background. For example: ``Our dashboard loads slowly (5-8 seconds) for users with large datasets. Profiling shows N+1 queries fetching user projects and statistics.'' This phase creates the foundational context that should remain relevant throughout the conversation.

\textbf{Phase 2: Divergent explorations} (3 topics, 3 turns each). The developer explores three alternative approaches that are intentionally unrelated to the core problem. These explorations introduce irrelevant context designed to test whether the LLM maintains focus. For the performance scenario, explorations include: comparing React vs Vue for frontend rewrites, discussing Next.js server-side rendering benefits, and evaluating TypeScript adoption. These topics represent realistic tangents—developers often investigate multiple solutions simultaneously—but are conceptually distant from the database query optimization problem.

\textbf{Phase 3: Synthesis query} (1 turn). The developer asks a follow-up question requiring focused attention on the original problem while introducing a new constraint. Continuing the performance example: ``Our users expect real-time updates, so caching is tricky. How can we optimize without making data stale?'' This query demands that the LLM recall the N+1 query problem and real-time requirement while ignoring the frontend framework discussions from Phase 2.

\subsubsection{Experimental Conditions}

We compare the following two conversation modes in the experimental setup:

\textbf{Linear (baseline):} All messages remain in a single conversation thread. The context for the final query includes: initial problem establishment (Phase 1), all three exploratory discussions (Phase 2), and the final query (Phase 3). Context size averages 31.0 messages. This represents current standard practice in conversational AI tools.

\textbf{Branched (intervention):} The developer creates a checkpoint after Phase 1, explores each alternative on a separate branch (3 branches total), then switches back to the main branch for Phase 3. The context for the final query includes only the initial problem establishment (Phase 1) and the final query. Exploratory discussions remain isolated on their respective branches. Context size averages 13.0 messages—a 58.1\% reduction.

\subsubsection{Evaluation Procedure}

For each scenario, we generate responses to the final synthesis query in both modes using GPT-4 (gpt-4-0125-preview). An independent LLM judge (also GPT-4, in a separate session with no knowledge of experimental conditions) evaluates response quality on five criteria:

\begin{enumerate}[leftmargin=*, itemsep=1pt]
\item \textbf{Relevance}: How directly the response addresses the specific question asked
\item \textbf{Accuracy}: Technical correctness of the advice provided
\item \textbf{Focus}: Clarity and lack of distraction from irrelevant topics
\item \textbf{Context Awareness}: Appropriate use of relevant prior discussion
\item \textbf{Specificity}: Concreteness of advice versus generic recommendations
\end{enumerate}

Each criterion receives a score from 1 (poor) to 10 (excellent). The judge is blind to the experimental condition and evaluates both responses in randomized order. This methodology follows established practices for LLM evaluation~\cite{liu2024lost,liang2024using}.

\subsubsection{Statistical Analysis}

We use paired t-tests to compare linear and branched modes within each scenario, controlling for scenario difficulty and content variation. Effect sizes are computed using Cohen's $d$ for paired samples. Given 30 scenarios, our design provides 80\% power to detect medium effects ($d \geq 0.5$) at $\alpha = 0.05$.

\subsection{Results}

\subsubsection{Overall Response Quality}

Table~\ref{tab:results} summarizes response quality across experimental conditions. Branched conversations achieved significantly higher overall scores than linear conversations ($8.67 \pm 0.39$ vs. $8.46 \pm 0.50$, $t(29)=2.73$, $p=0.010$, Cohen's $d=0.73$). This represents a 2.5\% improvement in response quality when context remains focused on relevant history.

\begin{table}[ht]
\centering
\caption{Response Quality: Linear vs Branched Conversations}
\label{tab:results}
\begin{threeparttable}
\begin{tabular}{lccccc}
\toprule
\textbf{Criterion} & \multicolumn{2}{c}{\textbf{Linear}} & \multicolumn{2}{c}{\textbf{Branched}} & \textbf{Improve}\\
& Mean & SD & Mean & SD & (\%) \\
\midrule
Relevance & 8.93 & 0.45 & 9.00 & 0.37 & +0.7\% \\
Accuracy & 9.03 & 0.32 & 9.07 & 0.45 & +0.4\% \\
Focus & 8.70 & 0.75 & 9.10 & 0.55 & +4.6\%** \\
Context Awareness & 7.37 & 0.85 & 7.87 & 0.73 & +6.8\%** \\
Specificity & 8.27 & 0.78 & 8.33 & 0.71 & +0.8\% \\
\midrule
\textbf{Overall} & 8.46 & 0.50 & 8.67 & 0.39 & +2.5\%*  \\
\bottomrule
\end{tabular}
\begin{tablenotes}
\small
\item All tests are paired t-tests ($n=30$). (**$p < 0.01$, *$p < 0.05$). 
\end{tablenotes}
\end{threeparttable}
\end{table}

\subsubsection{Criterion-Level Analysis}

\begin{figure}[t]
\centering
\includegraphics[width=\columnwidth]{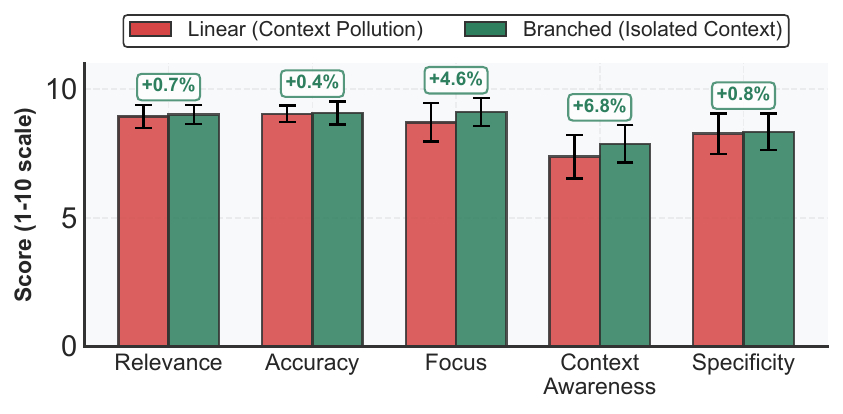}
\caption{Response quality improvements by evaluation criterion. Branched conversations showed large effect sizes for Focus (+4.6\%, d=0.80) and Context Awareness (+6.8\%, d=0.87) with significance p<0.01, validating that branching primarily improves context management rather than general reasoning.}
\Description{Bar graph showing comparision of response qualities improvements by evaluation criterion}
\label{fig:criteria}
\end{figure}

The largest improvements occurred in \emph{focus} ($+4.6\%$, $t(29)=3.52$, $p=0.001$, $d=0.80$) and \emph{context awareness} ($+6.8\%$, $t(29)=3.67$, $p=0.001$, $d=0.87$), as shown in Figure ~\ref{fig:criteria}. Both exhibit large effect sizes according to conventional thresholds ($d \geq 0.8$)~\cite{cohen1988statistical}, indicating that branched conversations enable substantially more focused responses that appropriately leverage relevant prior context without distraction from exploratory tangents.

\begin{figure}[t]
\centering
\includegraphics[width=0.9\columnwidth]{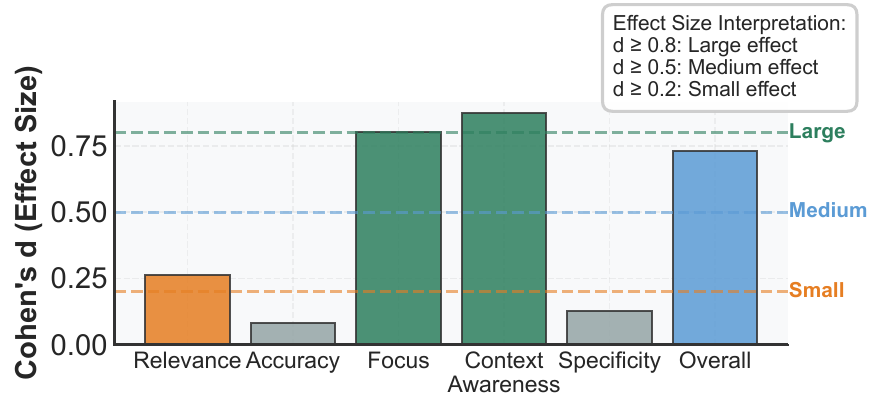}
\caption{Effect sizes (Cohen's d) for each evaluation criterion. Focus (d=0.80) and Context Awareness (d=0.87) exhibit large effect sizes, while Relevance (d=0.26), Accuracy (d=0.08), and Specificity (d=0.13) show small or negligible effects. Dashed lines indicate conventional effect size thresholds (small=0.2, medium=0.5, large=0.8). Overall effect size is d=0.73 (medium-large).}
\Description{Bar graph showing effect sizes for each evaluation criterion}
\label{fig:effect-sizes}
\end{figure}

Figure~\ref{fig:effect-sizes} provides a visual representation of effect 
sizes across all criteria. The large effects for focus and context awareness 
stand in contrast to the negligible-to-small effects for accuracy (d=0.08), 
specificity (d=0.13), and relevance (d=0.26). This pattern validates our 
hypothesis that branching addresses context management specifically rather 
than enhancing the LLM's general reasoning capabilities. The overall effect 
size (d=0.73) falls in the medium-to-large range, indicating a meaningful 
practical difference despite the modest 2.5\% absolute improvement.

Improvements in \emph{relevance} ($+0.7\%$, $d=0.26$), \emph{accuracy} ($+0.4\%$, $d=0.08$), and \emph{specificity} ($+0.8\%$, $d=0.13$) were smaller and not statistically significant individually, though they contributed to the overall improvement. The modest effects on relevance and accuracy suggest that branching primarily addresses context management rather than improving the LLM's fundamental reasoning capabilities.

\subsubsection{Context Efficiency}

\begin{figure}[t]
\centering
\includegraphics[width=0.85\columnwidth]{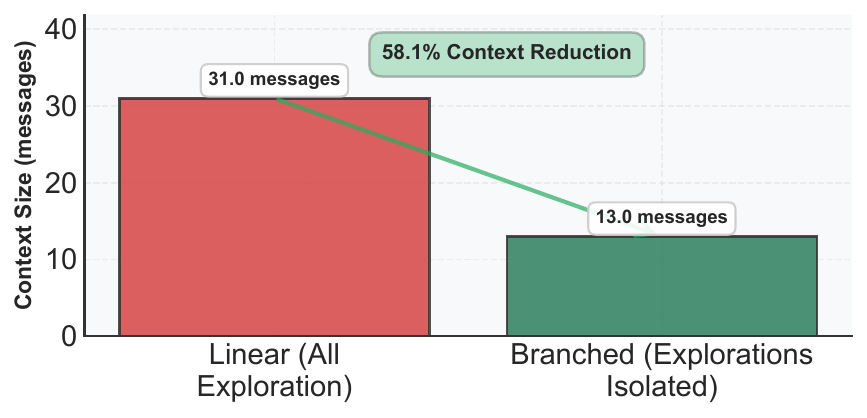}
\caption{Context size comparison between linear and branched conversations. Branched conversations maintained 58.1\% smaller context windows (13.0 vs 31.0 messages), eliminating the 18 intentionally polluting exploration messages. This reduction translates directly to lower API costs, reduced latency, and decreased risk of context-related errors.}
\Description{Graph showing context size reduction by 58.1\% using branched conversations}
\label{fig:context}
\end{figure}

Branched conversations maintained dramatically smaller context windows, as illustrated in Figure~\ref{fig:context}. Linear mode averaged 31.0 messages in context for the final query (the complete conversation history including all explorations). Branched mode averaged 13.0 messages (only the initial context establishment), representing a 58.1\% reduction. This reduction directly eliminates the 18 messages of intentionally polluting explorations from Phase 2, demonstrating that branch isolation successfully prevents irrelevant context accumulation.

\subsubsection{When Does Branching Help?}

Among the 30 scenarios, 20 (67\%) showed improvement with branching, 9 (30\%) showed no measurable change, and 1 (3\%) showed a small decrease. The scenarios with largest improvements (Scenarios 3 and 6, both $+13.2\%$, $+1.0$ points on the 10-point scale) involved exploring fundamentally conflicting architectural decisions where linear discussion caused the LLM to conflate approaches.

We analyzed the characteristics distinguishing high-benefit from zero-improvement scenarios. 
\newline
\textbf{High-benefit scenarios} (improvement $> 5\%$, $n=6$) shared three properties:

\begin{itemize}[leftmargin=*, itemsep=1pt]
\item \textbf{Conceptual distance}: Exploratory topics belonged to entirely different technical domains than the core problem (e.g., discussing frontend frameworks while solving database query optimization)
\item \textbf{Conflicting approaches}: Alternatives involved mutually exclusive architectural decisions (e.g., monolith versus microservices decomposition)
\item \textbf{Complex constraints}: Final queries introduced nuanced constraints requiring focused attention on original problem context (e.g., ``without adding latency'' or ``maintaining backward compatibility'')
\end{itemize}
\textbf{Zero-improvement scenarios} ($n=9$) typically involved:

\begin{itemize}[leftmargin=*, itemsep=1pt]
\item \textbf{Simple final queries}: Questions answerable independently of prior conversational context
\item \textbf{Low semantic overlap}: Exploratory topics sufficiently distinct that even GPT-4 in linear mode naturally ignored them
\item \textbf{Explicit constraint restatement}: Final queries that repeated all necessary context, making prior history less critical
\end{itemize}

\textbf{Illustrative example.} Consider Scenario 3 (Code Refactoring, $+13.2\%$ improvement). The developer establishes that their Rails UserController has grown to 600 lines with business logic mixed into HTTP concerns, making testing difficult. A small team needs a pragmatic refactoring approach. In linear mode, the developer then explores three unrelated topics: migrating from MySQL to PostgreSQL for JSON support, implementing WebSocket notifications, and configuring CDN for static assets. When the final query asks ``Given our small team new to design patterns, what's the most pragmatic first step that won't require a complete rewrite?'', the linear condition's response discussed database migration strategies and real-time notification architecture—topics from the polluting explorations—alongside service object patterns. The branched condition, with clean context focused solely on the 600-line controller problem, provided targeted advice on extracting command objects and form objects incrementally, directly addressing the ``small team'' and ``pragmatic'' constraints without distraction.

This analysis reveals that branching provides greatest value when: (1) developers explore conceptually distant alternatives, (2) final queries require nuanced understanding of original constraints, and (3) the problem space is complex enough that pollution could plausibly interfere. For simple queries or when alternatives are clearly separated by topic, linear conversations may suffice.

\subsection{Discussion}

Our results provide empirical evidence that conversation branching mitigates context pollution in exploratory programming tasks. The 2.5\% overall improvement, while modest in absolute terms, represents a medium-to-large effect size ($d=0.73$) and masks substantial benefits (up to 13.2\%) in scenarios involving conceptually distant explorations. The concentration of benefits in high-complexity scenarios validates that branching addresses a specific problem—context pollution from exploratory tangents—rather than universally improving all conversations.

The large effect sizes for focus ($d=0.80$) and context awareness ($d=0.87$) demonstrate that branching's primary benefit is maintaining clean context rather than enhancing the LLM's inherent capabilities. This aligns with our hypothesis: by isolating exploratory discussions on separate branches, the main conversation thread avoids accumulating irrelevant information that could bias subsequent responses.

The 58.1\% reduction in context size has practical implications beyond quality. Smaller contexts reduce API costs (fewer input tokens), improve latency (faster processing), and better align with documented LLM limitations in long-context scenarios~\cite{liu2024lost}. Our experimental design—scenarios with intentionally polluting explorations—demonstrates that this reduction eliminates genuinely distracting information rather than arbitrary message pruning. The correspondence between 58\% context reduction (eliminating 18 polluting messages) and improved focus scores provides evidence that isolation successfully prevents contamination.

\subsubsection{Design Implications}

These findings inform when to recommend branching in conversational AI interfaces. Systems could detect scenarios likely to benefit from branching through:

\begin{itemize}[leftmargin=*, itemsep=1pt]
\item \textbf{Semantic distance detection}: Compute embedding similarity between current topic and established problem context. When similarity drops below a threshold, suggest branching.
\item \textbf{Conversation pattern recognition}: Identify patterns like ``What if we tried [alternative approach]?'' that signal exploratory divergence.
\item \textbf{Complexity heuristics}: For architecturally significant decisions or multi-constraint problems, proactively recommend checkpointing before exploration.
\end{itemize}

Conversely, for simple queries, closely related explorations (e.g., comparing two similar libraries), or when developers explicitly restate all context in their queries, branching overhead may not be justified.

\subsubsection{Threats to Validity} While every effort was made to ensure the rigor of this study, several threats to validity may influence the interpretation of the results.

\textbf{LLM-as-judge methodology.} Our evaluation uses GPT-4 to judge GPT-4-generated responses, which may introduce bias despite blinding to experimental conditions. However, our ground truth annotations—explicit lists of concepts that should versus should not influence responses—enable verification that improvements stem from maintaining correct focus. Manual inspection of high-improvement scenarios confirms that branched responses discussed topics from the ``should remember'' list while avoiding topics from the ``should not be distracted by'' list at higher rates than linear responses.

\textbf{Scenario design.} Our scenarios deliberately inject irrelevant explorations to test context pollution. While this represents realistic exploratory programming patterns—developers naturally investigate multiple alternatives, some of which prove unrelated—the explorations may be more obviously irrelevant than in organic workflows. However, this conservative design strengthens our claims: if branching improves focus even when pollution is relatively obvious, benefits should persist when distractions are more subtle.

\textbf{Single model evaluation.} Our experiments used GPT-4 for both generating responses and evaluating them. While GPT-4 represents current state-of-the-art, the multi-turn degradation problem documented by Laban et al.~\cite{laban2025lost} affects all tested LLMs (39\% average across 15 models). We expect branching to provide similar benefits with other models, though the magnitude may vary.

\textbf{Generalization.} The scenarios, while realistic, are synthetic rather than drawn from actual developer workflows. The standardized structure (5 initial turns, 3 explorations, 1 final query) may not reflect the diversity of real exploratory programming sessions. Future work should validate these findings with developers using \system{} for actual programming tasks, measuring both response quality and subjective usability.

\subsubsection{Comparison to Related Work}

Our 2.5\% improvement may appear modest compared to the 39\% multi-turn degradation documented by Laban et al.~\cite{laban2025lost}. However, their study measured degradation across entire conversations (up to 10 turns), while our intervention targets a specific failure mode: context pollution from exploratory tangents. Branching addresses one component of multi-turn challenges rather than solving the entire problem. Complementary approaches—better prompting, retrieval-augmented generation, model improvements—remain necessary for comprehensive solutions.

Our results align with findings from Vaithilingam et al.~\cite{vaithilingam2022expectation}, who documented that developers spend significant effort managing conversation context with AI coding assistants. By providing explicit primitives for context isolation, \system{} addresses a documented pain point in AI-assisted programming workflows.

\section{Conclusion}

We presented \system{}, a conversation management system that applies version control semantics to LLM interactions. Through four core primitives—checkpoint, branch, switch, and inject—\system{} enables developers to explore alternative approaches in isolation without polluting the main conversational thread. Our controlled experiment with 30 software engineering scenarios demonstrated that branching improves response quality by 2.5\% overall (p=0.010, Cohen's d=0.73), with large effect sizes for focus (d=0.80) and context awareness (d=0.87). These targeted improvements validate our hypothesis: isolating exploratory discussions prevents context pollution that degrades LLM responses.

Benefits were concentrated in complex scenarios involving conceptually distant explorations (peak improvement 13.2\%), while simple queries showed minimal benefit (30\% of scenarios). This heterogeneity reveals when branching addresses a genuine problem versus when linear conversations suffice. Simultaneously, branching reduced context size by 58.1\%, eliminating irrelevant exploratory content and providing practical benefits beyond quality: lower API costs, reduced latency, and better alignment with documented LLM context limitations.

Our work establishes conversation branching as a fundamental primitive for AI-assisted exploratory work. Just as version control enables developers to explore code alternatives through isolated branches and selective merging, \system{} enables exploration of conversational alternatives while maintaining deterministic, reproducible context states. Beyond software engineering, this primitive applies to any domain involving exploratory interaction with AI systems: research workflows investigating alternative analyses, creative writing exploring narrative branches, or educational interactions allowing tangential questions without losing the main thread.

\subsection{Future Work}

Our work opens several directions for future research:

\textbf{Automatic branch suggestion.} Current \system{} requires manual checkpoint and branch creation. Machine learning models could detect conversational patterns indicating exploratory divergence (``What if we tried...?'', semantic distance from established context) and proactively suggest branching. Our finding that benefits concentrate in scenarios with conceptually distant explorations provides a starting point for such heuristics.

\textbf{Intelligent injection.} Manual message selection for injection is tedious. Future work could develop techniques to automatically identify valuable insights from exploratory branches—perhaps using embedding similarity to the main branch's context or tracking which explorations the user references in subsequent queries.

\textbf{Integration with development environments.} AI coding assistants like GitHub Copilot and Cursor are embedded in IDEs. Conversation branching should integrate at this level, with branches tied to code branches, checkpoints synchronized with git commits, and exploratory conversations linked to specific files or functions under development.

\textbf{Long-term conversation management.} Our evaluation focused on single-session scenarios. Real development projects span days or weeks. Future work should explore how branching interacts with conversation persistence, how to manage dozens of branches accumulated over time, and whether branch pruning strategies from Git (garbage collection, branch deletion) apply to conversations.


\textbf{Model-specific branching.} Future agent systems could exploit the complementary strengths of different LLMs by routing subtasks to specialized models. A primary reasoning trunk may handle high-level planning, while separate branches perform code generation or research synthesis. Such model-aware branching can improve efficiency and overall task performance in complex workflows.

We release \system{} as open source \footnote{Available at: https://github.com/glanzz/context-branching} along with our experimental scenarios, evaluation framework, and analysis scripts. We invite the research community to build upon this foundation, exploring how conversation branching can enhance human-AI collaboration across domains.

\subsection{Closing Perspective}

The history of human-computer interaction is marked by the introduction of fundamental interaction primitives that, once established, become invisible infrastructure. The mouse enabled direct manipulation. Copy-paste enabled information reuse. Undo/redo enabled fearless experimentation. Version control enabled collaborative exploration of alternatives. We believe conversation branching will similarly become fundamental infrastructure for AI-assisted work—not because it produces dramatic performance gains in all cases, but because it provides the right abstraction for a common problem: managing multiple lines of inquiry without mutual interference.

Our 2.5\% average improvement may seem modest, but it represents validation of a mechanism that addresses a documented limitation of current LLM interactions. As models improve and multi-turn degradation potentially decreases, the relative benefit of branching may change. However, the fundamental problem—wanting to explore alternatives without committing to a single conversational path—will persist. Conversation branching provides the primitives to support this exploration, just as version control provides primitives to support code exploration.

The future of AI-assisted work lies not just in more capable models, but in better interaction paradigms that align with how humans naturally think and work. Conversation branching is one such paradigm, and we look forward to seeing how the community builds upon this foundation.
\balance

\bibliographystyle{ACM-Reference-Format}
\bibliography{references}

\end{document}